\documentstyle[12pt]{article}

\begin{document}

\mbox{} \hskip 10cm DF/IST-3.2000

\begin{center}
{\bf False vacuum decay: effective one-loop action for 
pair creation of domain walls} \\
\vskip 1cm
\'Oscar J. C. Dias \\
\vskip 0.3cm
{\scriptsize  CENTRA, Departamento de F\'{\i}sica,
      Instituto Superior T\'ecnico,} \\
{\scriptsize  Av. Rovisco Pais 1, 1049-001 Lisboa, Portugal.} \\
{\scriptsize  oscar@fisica.ist.utl.pt} \\
\vskip 0.6cm
             Jos\'e P. S. Lemos   \\
\vskip 0.3cm
{\scriptsize  CENTRA, Departamento de F\'{\i}sica,
      Instituto Superior T\'ecnico,} \\
{\scriptsize  Av. Rovisco Pais 1, 1049-001, Lisboa, Portugal.} \\
{\scriptsize  lemos@kelvin.ist.utl.pt}
 \end{center}
\bigskip


\begin{abstract}
\noindent

An effective one-loop action built from the soliton field itself for
the two-dimensional (2D) problem of soliton pair creation is
proposed. The action consists of the usual mass term and a kinetic
term in which the simple derivative of the soliton field is replaced
by a covariant derivative. In this effective action the soliton 
charge is treated no longer as a topological charge but as a 
Noether charge. Using this effective one-loop action, the
soliton-antisoliton pair production rate $\Gamma/L=A\:\exp{[-S_0]}$
is calculated and one recovers Stone's exponential factor $S_0$
and the prefactor $A$ of Kiselev, Selivanov and Voloshin.
The results are also valid straightforwardly to the problem of pair
creation rate of domain walls in dimensions D$\geq$3.
\strut  
\end{abstract}
\newpage

\noindent
{\bf I. Introduction}
\vskip 3mm

Stone \cite{Stone} has studied the problem of a scalar field theory in
(1+1)D with a metastable vacuum, i.e., with a scalar potential that has
a false vacuum, $\phi_+$, and a true vacuum, $\phi_-$, separated by an
energy density difference, $\epsilon$. Stone has noticed that the
decay process can be interpreted as the false vacuum decaying into the
true vacuum plus a creation of a soliton-antisoliton pair: $\phi_+
\rightarrow \phi_- +s +\bar{s}\:.$ The energy necessary for the
materialization of the pair comes from the energy density difference
between the two vacua. The soliton-antisoliton pair production rate
per unit time and length, $\Gamma/L$, can then be identified with the
decay rate of the false vacuum and is given by ($\hbar=c=1$)
\cite{Stone}:
\begin{eqnarray} 
\Gamma/L=A\:{\rm e}^{-S_0}=A\:{\rm e}^{-\frac{\pi m^2}{\epsilon}}\:,  
\label{0.1}
\end{eqnarray}
where $m$ is the soliton mass and prefactor $A$ is a 
functional determinant whose value was first calculated by Kiselev 
and Selivanov \cite{Kiselev1,Kiselev2} and later by Voloshin  
\cite{Voloshin1}. Extensions to this decay problem, such as 
induced false vacuum decay, have been studied by several authors 
(for a review and references see, e.g., \cite{Voloshin3,Kiselev3}). 

The method used in \cite{Stone,Kiselev1,Kiselev2,Voloshin1} is based
one the instanton method introduced by Langer in his work about decay
of metastable termodynamical states \cite{Langer}.  This powerful
method has been applied to several different studies, namely: Coleman
and Callan \cite{Coleman,ColemanCallan} have computed the bubble
production rate that accompanies the cosmological phase transitions in
a (3+1)D scalar theory (this was indeed previously calculated by other
methods by Voloshin, Kobzarev and Okun \cite{Voloshin2}); Affleck and
Manton \cite{Affleck2} have studied monopole pair production in a weak
external magnetic field and Affleck, Alvarez and Manton
\cite{Affleck1}, have studied $e^+e^-$ boson pair production in a weak
external electric field. Recent developments studying pair production
of boson and spinorial particles in external Maxwell's fields have
been performed by several authors using different methods
\cite{gavri}-\cite{lin} and  similar results in 
the Euler-Heisenberg theory, a modified Maxwell theory, have been also
obtained \cite{KS}. The decay of false vacuum in a condensed matter system
providing soliton tunneling has been studied in \cite{Miller,Miller2}.

In this paper we propose an effective one-loop action
built from the soliton field itself to study the problem of
Stone \cite{Stone},  Kiselev and Selivanov 
\cite{Kiselev1,Kiselev2} and Voloshin \cite{Voloshin1}. 
The action consists of the usual mass term
and a kinetic term in which the simple derivative of the 
soliton field is replaced by a kind of covariant 
derivative.  In this effective action the soliton 
charge is treated no longer as a topological charge but as a 
Noether charge. This procedure of working with an effective 
action for the soliton field itself has been introduced by Coleman 
 \cite{Coleman1} where the equivalence between the 
Sine-Gordon model and the Thirring model was shown, and by 
Montonen and Olive
\cite{Olive} who  have proposed an equivalent dual field theory
for the Prasad-Sommerfield monopole soliton.
More connected to our problem, Manton \cite{Manton} has proposed an
effective action built from the soliton field  itself 
which reproduces the soliton physical properties of 
(1+1)D nonlinear scalar field theories having 
symmetric potentials with degenerate minima. In this 
paper we deal instead with a potential with non-degenerate minima 
in a (1+1)D scalar field theory. Thus, our effective 
action is new since Manton was not dealing with 
the soliton pair production process.

Using the effective one-loop action and the method  
presented in \cite{Affleck1}, we calculate the 
soliton-antisoliton pair production rate, (\ref{0.1}).  
One recovers Stone's exponential factor $S_0$ \cite{Stone}
and the prefactor $A$ of Kiselev and Selivanov 
\cite{Kiselev1,Kiselev2} and Voloshin \cite{Voloshin1}.

\vskip 0.5cm

\noindent
{\bf II. Effective one-loop action}

\vskip 3mm

In order to present some useful soliton properties 
let us consider 
a scalar field theory in a (1+1)D  spacetime, 
whose dynamics is governed by the action (see, e.g., 
\cite{Rajamaran}),
\begin{equation}
S[\phi(x,t)]=\int d^2x {\biggl [} \frac{1}{2}
\partial_{\mu}\phi \partial^{\mu}\phi - 
U(\phi) {\biggl ]},
\label{1.1}
\end{equation}
where $U$ is a generic potential. A particular important case
is when $U$ is a symmetric potential, $U=U_{\rm s}(\phi)$,
with two or 
more degenerate minima. In the $\phi^4$ theory  the 
potential is $U_{\rm s}(\phi)=\frac{1}{4}\lambda
{\bigl(}\phi^2-\mu^2/\lambda{\bigr)}^2$, with 
$\mu\geq0$ and $\lambda \geq0\:$. Stationarizing the action
one obtains the solutions of the theory which have finite 
and localized energy. The solutions are the soliton
\begin{equation} 
\psi\equiv\phi_{\rm sol}=+
\frac{\mu}{\sqrt{\lambda}}\tanh{\biggl[}
\frac{\mu}{\sqrt{2}} \:\frac{(x-x_{0})-vt}{\sqrt{1-v^2}}
{\biggr]}\:,  
\label{1.2}
\end{equation}
and the antisoliton $-\psi$.
From the hamiltonian density, 
${\cal H}=\frac{1}{2}(\partial_{x}\phi)^2 + 
U_{\rm s}(\phi)$, one can calculate the mass 
of the soliton and antisoliton 
\begin{equation}
m=\int_{-\infty}^{+\infty} dx
\,{\cal H}(x)=\frac{2\sqrt{2}}{3}
\frac{\mu^3}{\lambda}.
\label{1.3}
\end{equation} 
One can also define the topological charge, 
$Q=\frac{1}{2}
{\bigl[}\psi(x=+\infty)-\psi(x=-\infty){\bigr]}$, 
(conserved in time) which has the positive value
$Q_{\rm s}=+\mu/ \sqrt{\lambda}$ in the case of the soliton
and the negative value
$Q_{\bar{\rm s}}=-\mu/ \sqrt{\lambda}$ in the case of the antisoliton.
To this charge one associates the topological current
$k^{\mu}=\frac{1}{2} 
\varepsilon^{\mu\nu} \partial_{\nu} \psi$ which is 
conserved, $\partial_{\mu}k^{\mu}=0$, and such 
that $Q=\int_{-\infty}^{+\infty} dx
k^0$. 

Now, let us consider a non-degenerate potential $U$ in 
action (\ref{1.1}) by adding to $U_{\rm s}$
 a small term that breaks its symmetry 
\cite{Stone,ColemanCallan}:
$U(\phi)=U_{\rm s}(\phi)+\frac{\epsilon}{2\mu/ 
\sqrt{\lambda}}(\phi-\mu/ \sqrt{\lambda})$, 
where $\epsilon$ is the energy density (per unit length) 
difference between the true 
($\phi_-=-\mu/ \sqrt{\lambda}$) and false 
($\phi_+=+\mu/ \sqrt{\lambda}$) vacua. 
As noticed in \cite{Stone,Kiselev1,Kiselev2}, $\epsilon$ is 
responsible for both the decay of false vacuum and 
soliton-antisoliton pair creation.

We want to find an effective one-loop action built from
the soliton field itself and that describes the above 
pair creation process. The soliton field should be a 
charged field since the system admits  two charges, 
$Q_{\rm s}$ and $Q_{\bar{\rm s}}$. Therefore, the action 
should contain the mass term $m^2\psi \psi^*$, where $m$ 
is the soliton mass given in eq.(\ref{1.3}), and the 
kinetic term $(\partial_{\mu}\psi)(\partial^{\mu}\psi^*)$.
Thus, the free field effective action is $\int d^2x
[(\partial_{\mu}\psi)(\partial^{\mu}\psi^*)-
m^2\psi \psi^*]$. However, if one demands local gauge 
invariance one has to introduce an ``electromagnetic''
2-vector potential $A_{\mu}$ 
which transforms the common derivative 
$\partial_\mu \psi$ into a covariant derivative 
$(\partial_\mu +i Q_{\rm s}A_{\mu})\psi$.
As is well known, the field $A_{\mu}$ itself should 
contribute to the action. This contribution must be gauge
invariant since the covariant kinetic term
plus the mass term are already gauge invariant. This 
is achieved by defining the invariant 2-form field,
$F_{\mu\nu}=\partial_\mu A_{\nu}-\partial_\nu A_{\mu}$. In 
two dimensions, an anti-symmetric field can only be of the 
form: $F_{\mu\nu}=\sigma(t,x) \varepsilon_{\mu\nu}$, where
$\varepsilon_{\mu\nu}$ is the Levi-Cevita tensor and  
$\sigma(t,x)$ a function. Therefore the effective one-loop
action should be $S^{\rm eff}=\int d^2x
[(\partial_{\nu}\psi+i Q_{\rm s}A_{\nu}\psi)(\partial^{\nu}\psi^*
-i Q_{\rm s}A^{\nu}\psi^*)-
m^2\psi \psi^*-\frac{1}{4}F_{\mu\nu}F^{\mu\nu}]$.

Note now that the charged soliton acts also as a source, 
thus modifying the surrounding field. As a first 
approximation we shall neglect this effect and assume 
$A_\mu$ fixed by external conditions. This allows us 
to drop the contribution of the term 
$F_{\mu\nu}F^{\mu\nu}$ in the effective action. 
Moreover, the external field responsable for the 
pair creation is essentially represented by the 
energy density difference $\epsilon$ so we postulate 
that $F^{\rm ext}_{\mu\nu}= \frac{\epsilon}{\mu/ 
\sqrt{\lambda}} \varepsilon_{\mu\nu}$. Therefore,
$A^{\rm ext}_{\mu}$ is given by 
$A^{\rm ext}_{\mu}=\frac{1}{2}\frac{\epsilon}{\mu/ 
\sqrt{\lambda}} \varepsilon_{\mu\nu}x^{\nu}$.

Finally, if the system is analytically continued  to 
euclidean spacetime $(t_{\rm Min} \rightarrow -it_{\rm Euc}$; 
$A_0 \rightarrow iA_2)$ one obtains the euclidean effective 
one-loop action for the soliton pair creation problem 
\begin{eqnarray} 
S^{\rm eff}_{\rm Euc}=\int d^2x
{\biggl [} {\bigl |}(\partial_{\mu}-\frac{1}{2} \epsilon
\:\varepsilon_{\mu\nu}x_\nu)\psi{\bigl |}^2+m^2|\psi|^2 
{\biggr ]}\:.
\label{1.4}
\end{eqnarray} 

In the next section this euclidean effective 
one-loop action is going to be used to calculate the 
soliton-antisoliton pair production rate (\ref{0.1}). 
Although the calculations
are now similar to those found in Affleck, Alvarez and Manton 
pair creation problem \cite{Affleck1}, we present some 
important steps and results since in two dimensions they are 
slightly different.

\vskip 0.5cm

\noindent
{\bf III. Pair production rate}

\vskip 3mm

The soliton-antisoliton pair production rate per unit
time is equal to the false vacuum decay rate per unit time 
\begin{eqnarray}
\Gamma=- 2\:{\rm Im}E_0\:, \label{2.1}
\end{eqnarray} 
where the vacuum energy, $E_0$, is given by 
the euclidean functional integral 
\begin{eqnarray} 
{\rm e}^{-E_0 T} = \lim_{T
\rightarrow \infty} \int [{\cal D}\psi] [{\cal D}\psi^*] 
{\rm e}^{-S^{\rm eff}_{\rm Euc}(\psi;\psi^*)
}\:. \label{2.2}
\end{eqnarray} 
As it will be verified,
$E_0$ will receive a small imaginary contribution
from the negative-mode associated to the quantum 
fluctuations about the instanton (which 
stationarizies the action) and this fact is 
responsible for the decay. Combining (\ref{2.1}) 
and (\ref{2.2}) one has
\begin{eqnarray}
\Gamma
= \lim_{T \rightarrow \infty} \frac{2}{T} 
{\rm Im}\ln \int [{\cal D}\psi] [{\cal D}\psi^*] 
{\rm e}^{-S^{\rm eff}_{\rm Euc}(\psi;\psi^*)}\:,
 \label{2.3}
\end{eqnarray} 
where $S^{\rm eff}_{\rm Euc}$ is given by (\ref{1.4}). 
Integrating out $\psi$ and $\psi^*$ in (\ref{2.3}) one obtains
\begin{eqnarray} 
\Gamma=-\lim_{T \rightarrow \infty}
\frac{2}{T}\:{\rm Im}\:{\rm tr}\:\ln{\bigl
[}(\partial_{\mu}-\frac{1}{2} \epsilon
\:\varepsilon_{\mu\nu}x_\nu)^2+ m^2 {\bigr ]}.  \label{2.5}
\end{eqnarray} 
The logarithm in (\ref{2.5}) can be written as a ``Schwinger proper time 
integral", $\ln u=-\int_{0}^{\infty}\frac{d \cal{T}}{\cal{T}} \exp{\bigl
(}-\frac{1}{2}u \cal{T}{\bigr )}$. Taking $u={\bigl
[}(\partial_{\mu}-\frac{1}{2} \epsilon
\:\varepsilon_{\mu\nu}x_\nu)^2+ m^2 {\bigr ]}$, yields
\begin{eqnarray} 
\Gamma=\lim_{T \rightarrow \infty}
\frac{2}{T}\:{\rm Im}\int_{0}^{\infty}\frac{d \cal{T}}{\cal{T}}
{\rm e}^{-\frac{1}{2}m^2 \cal{T}} \:{\rm tr}\:\exp{\biggl [}-\frac{1}{2}{\biggl
(}P_\mu-\frac{1}{2} \epsilon \:\varepsilon_{\mu\nu}x_\nu{\biggl
)}^2 \cal{T}{\biggr ]}. \label{2.6} 
\end{eqnarray} 
Notice that now the trace is of the form ${\rm tr}\:{\rm e}^{-H \cal{T}}$, with $
H=\frac{1}{2}{\biggl [}P_\mu-\frac{1}{2} \epsilon
\:\varepsilon_{\mu\nu}x_\nu{\biggr ]}^2$ being the Hamiltonian for
 a particle subjected to the interaction with the external
scalar field in a (2+1)D spacetime, and the proper time playing the
role of a time coordinate. One has started with a scalar field theory in
a euclidean 2D spacetime and now one has found an effective theory for
particles in a 3D spacetime. It is in this new context that the pair
production rate is going to be calculated. The gain in having the
trace in the given form is that it can be written as a path integral
${\rm tr}\:{\rm e}^{-H \cal{T}}=\int [dx] \exp{\biggl [}-\int d {\cal{T}} \:L 
{\biggr ]}$,
where $L=\frac{1}{2}\dot{x}_\mu\dot{x}_\mu+\frac{1}{2} \epsilon
\:\varepsilon_{\mu\nu}x_\nu\dot{x}_\mu$ is the Lagrangian associated
with our Hamiltonian. Thus, 
\begin{equation} 
\Gamma=\lim_{T \rightarrow \infty}
\frac{2}{T}\:{\rm Im}\int_{0}^{\infty}\frac{d \cal{T}}{\cal{T}}
{\rm e}^{-\frac{1}{2}m^2 \cal{T}} \:\int [dx] \exp{\biggl
[}\!\!-\!\!\int_{0}^{\cal{T}}\!\!d {\cal{T}}{\biggl (}
\frac{1}{2}\dot{x}_\mu\dot{x}_\mu+\frac{1}{2} \epsilon
\:\varepsilon_{\mu\nu}x_\nu\dot{x}_\mu{\biggr )} {\biggr ]}.
\label{2.7} 
\end{equation}
Rescaling the proper time variable, $d {\cal{T}} \rightarrow\frac{d\tau}
{\cal{T}}$, 
and noticing that the path integral is over all the paths,
$x_\mu(\tau)$, such that $x_\mu(1)=x_\mu(0)$, one has
\begin{equation}
\Gamma=\lim_{T \rightarrow \infty} \frac{2}{T}\:{\rm
Im}\int [dx]{\rm e}^{-\frac{1}{2} \epsilon \oint \varepsilon_{\mu\nu}x_\nu
dx_\mu}\! \int_{0}^{\infty}\frac{d {\cal{T}}}{\cal{T}}
\exp{\biggl [}\!\!-\!{\biggl
(}\frac{1}{2}m^2 {\cal{T}}+\frac{1}{2 \cal{T}}\!\int_{0}^{1}\!\!d\tau
\dot{x}_\mu\dot{x}_\mu{\biggr )}{\biggr ]}\!. 
\label{2.8} 
\end{equation}
The $\cal{T}$ integral can be calculated expanding the function about 
the stationary point ${\cal{T}}_0^2=\frac{\int_{0}^{1}
d\tau\dot{x}^2}{m^2}$:
\begin{eqnarray} 
\int\frac{d \cal{T}}{\cal{T}}{\rm e}^{-f(\cal{T})}\sim
{\rm e}^{-f({\cal{T}}_0)}\frac{1}{{\cal{T}}_0}\sqrt{\frac{\pi}
{\frac{1}{2}f''({\cal{T}}_0)}}\sim
 {\rm e}^{-m\sqrt{\int_{0}^{1}
d\tau\dot{x}^2}}\frac{1}{m}\sqrt{\frac{2\pi}{{\cal{T}}_0}}\:. 
\label{2.9} 
\end{eqnarray}
Then (\ref{2.8}) can be written as 
\begin{eqnarray} 
\Gamma = \lim_{T
\rightarrow \infty}
\frac{1}{T}\:\frac{2}{m}\sqrt{\frac{2\pi}{{\cal{T}}_0}}{\rm Im}\int
[dx]{\rm e}^{-S_{\rm Euc}[x_\mu(\tau)]} \:,
\label{2.10} 
\end{eqnarray}
where $S_{\rm Euc}=m\sqrt{\int_{0}^{1}
d\tau\dot{x}_\mu\dot{x}_\mu}+\frac{1}{2} \epsilon \oint
\varepsilon_{\mu\nu}x_\nu dx_\mu$.
This integral can be solved using the instanton method. Stationarizing
the action, one gets the equation of motion in the (2+1)D
spacetime 
\begin{eqnarray}
\frac{m
\ddot{x}_\mu(\tau')}{\sqrt{\int_{0}^{1} d\tau\dot{x}^2}}=-\epsilon
\:\varepsilon_{\mu\nu}\dot{x}_\nu(\tau')\:;\hspace{.1in}{\rm
with}\:\:\mu=1,2\:\:{\rm and}\:\:\dot{x}_\mu=\frac{d x_\mu}{d\tau}\:.
\label{2.11} 
\end{eqnarray}
The instanton, $x_\mu^{\rm cl}(\tau)$, i.e., the solution of the 
euclidean equation of motion that obeys the boundary conditions
$x_\mu(\tau=1)=x_\mu(\tau=0)$ is
\begin{eqnarray} 
x_\mu^{\rm cl}(\tau)=R(\cos2 \pi \tau, \sin 2 \pi
\tau)\:;\:\:\:\:\:{\rm with}\:\:\:R=\frac{m}{\epsilon}\:. 
\label{2.12} 
\end{eqnarray} 
The instanton represents a particle describing a loop
of radius $R$ in the plane defined by the time $x_2$ and 
by the direction $x_1$. The loop is
 a thin wall that separates the true vacuum located inside the loop
 from the false vacuum outside.

The euclidean action of the instanton is given by
$S_0=S_{\rm Euc}[x_\mu^{\rm cl}(\tau)]=
m 2 \pi R-\epsilon\pi R^2$. The first term is the rest 
energy of the particle times the orbital length and the 
second term represents the interaction of the particle 
with the external scalar field. The loop radius, $R=m/ \epsilon$, 
stationarizies the instanton action. 
The action is then $S_0=\pi m^2/ \epsilon$.  

The second order variation operator is given by
\begin{eqnarray}
\hspace{-.2in}M_{\mu\nu} \!\!\!\!&\equiv&\!\!\!\! \frac{\delta^2
S}{\delta x_\nu(\tau') \delta x_\mu(\tau)} {\biggr |}_{x^{\rm
cl}}=\nonumber \\ 
&=&\!\!\!\! {\biggl [}-{\biggl
(}\frac{m\delta_{\mu\nu}}{\sqrt{\int_{0}^{1}
d\tau\dot{x}^2}}\frac{d^2}{d\tau^2} +\epsilon
\:\varepsilon_{\mu\nu}\frac{d }{d\tau}{\biggr )}\delta(\tau-\tau')-
\frac{m \ddot{x}_{\mu}(\tau) \ddot{x}_{\nu}(\tau')}{{\bigl
[}\int_{0}^{1} d\tau\dot{x}^2{\bigr ]}^{3/2}} {\biggr ]}_{x^{\rm cl}}
=\nonumber \\
&=&\!\!\!\! -{\biggl
[}\frac{\epsilon}{2\pi}\delta_{\mu\nu}\frac{d^2}{d\tau^2} +\epsilon
\:\varepsilon_{\mu\nu}\frac{d }{d\tau}{\biggr ]}\delta(\tau-\tau')-
\frac{2\pi \epsilon x_{\mu}^{\rm cl}(\tau) x_{\nu}^{\rm
cl}(\tau')}{R^2}\:. 
\label{2.14} 
\end{eqnarray} 
The eigenvectors $\eta_\mu^n$, and the eigenvalues  $\lambda_n$,
associated with the operator $M_{\mu\nu}$ are such that 
\begin{equation}
M_{\mu\nu}\:\eta_\nu^n(\tau')=\lambda_n
\:\eta_\mu^n(\tau')\:\delta(\tau-\tau')\:.
\end{equation}
From this one concludes that:
\newline
${\bf{(i)}}$ the positive eigenmodes are: 
\newline
$(\cos 2n\pi \tau, \sin 2n\pi \tau)$ and
$(\sin 2n\pi \tau, -\cos 2n\pi \tau)$ with
$\lambda_n=2\pi \epsilon(n^2-n)$,  $n=2,3,4...$; 
\newline
$(\sin 2n\pi \tau, \cos 2n\pi \tau)$ and 
$(\cos 2n\pi \tau, -\sin 2n\pi \tau)$ with
$\lambda_n=2\pi \epsilon(n^2+n)$, $n=1,2,3...$;
\newline ${\bf{(ii)}}$ there are two zero-modes associated with the 
translation of the loop along the $x_1$ and $x_2$ directions: 
$(1, 0)$ and $(0, 1)$ with $\lambda=0$;
\newline ${\bf{(iii)}}$ there is a zero-mode associated with the 
translation along the proper time, $\tau$: 
$(\sin 2\pi \tau, -\cos 2\pi
\tau)=-\frac{\dot{x}_\mu^{\rm cl}}{2\pi R}$ with $\lambda=0$; 
\newline ${\bf{(iv)}}$ there is a single negative mode associated 
to the change of the loop radius $R$: $(\cos 2\pi \tau, \sin 2\pi
\tau)=\frac{x_\mu^{\rm cl}}{R}$  with $\lambda_{-}=-2\pi
\epsilon$.  

Now, we consider small fluctuations about the instanton, i.e., we do 
$x_\mu(\tau)=x_\mu^{\rm cl}(\tau)+\eta_\mu(\tau)$. 
The euclidean action is expanded to second order so that the 
path integral
(\ref{2.10}) can be aproximated by
\begin{equation} 
\Gamma
\simeq \lim_{T \rightarrow \infty}
\frac{1}{T}\frac{2}{m}\sqrt{\frac{2\pi}{{\cal{T}}_0}}\:
{\rm e}^{-S_0}\:{\rm
Im} \int [d\eta(\tau)]\exp {\biggl[}-\frac{1}{2}\int d\tau
d\tau'\,\eta_\mu(\tau)\: M_{\mu\nu} \:\eta_\nu(\tau'){\biggr]}.
\label{2.20} 
\end{equation} 
The path integral in equation (\ref{2.20}) is the one-loop factor and
is given by ${\cal N}{\bigl (}{\rm Det}M{\bigr )} ^{-\frac{1}{2}}
={\cal N} \prod \,(\lambda_n)^{-\frac{1}{2}}$, where $\lambda_n$ are
the eigenvalues of $M_{\mu\nu}$ and ${\cal N}$ is a normalization
factor that will not be needed. To overcome the problem that arises
from having an infinite product of eigenvalues, one compares our
system with the free particle system 
\begin{eqnarray} 
\int [d\eta]\exp {\biggl
[}\!\!\!\!\!\!&-&\!\!\!\!\frac{1}{2}\int\! d\tau
d\tau'\,\eta_\mu(\tau)\: M_{\mu\nu} \:\eta_\nu(\tau'){\biggr ]}
= \nonumber \\
&=& 
\int [d\eta]\exp \!{\biggl [}\!\!-\frac{1}{2}\int\! d\tau
d\tau'\,\eta_\mu(\tau)\:M^0_{\mu\nu} \:\eta_\nu(\tau'){\biggr ]}
\frac{\prod \,(\lambda_n)^{-\frac{1}{2}}}{\prod
\,(\lambda'_n)^{-\frac{1}{2}}}\:,  
\label{2.21} 
\end{eqnarray} 
where $M^0_{\mu\nu} =-\frac{1}{{\cal{T}}_0}
\delta_{\mu\nu}\frac{d^2}{d\tau^2} \delta(\tau-\tau')$ 
is the second variation operator of the free system with eigenvalues 
$\lambda'_n =2 \pi \epsilon n^2\:,\:\:\:\:\:n=0,1,2,3...$ 
(each with multiplicity 4). In equation (\ref{2.21}) the first factor 
is the path integral of a free particle in a (2+1)D euclidean spacetime
\begin{equation} 
\int [d\eta]\exp \!{\biggl [}\!-\frac{1}{2}\int\!
d\tau d\tau'\,\eta_\mu\:M^0_{\mu\nu} \:\eta_\nu{\biggr
]}=\int [d\eta]\exp \!{\biggl [}\!-\frac{1}{2{\cal{T}}_0}\int\! d\tau
\,\dot{\eta}_\mu\:\dot{\eta}_\mu{\biggr ]}=
\frac{1}{2\pi {\cal{T}}_0} \,.
\label{2.22} 
\end{equation}
In the productory, one omits the zero eigenvalues, but one has to 
introduce the normalization factor $\frac{||dx^{\rm cl}_\mu/d
\tau||}{||\eta^0_\mu||}\sqrt{\frac{1}{ 2\pi}}=\sqrt{2 \pi}R$ 
 which is  associated with the proper time eigenvalue.
In addition, associated with the negative eigenvalue one has to 
introduce a factor of $1/2$ which accounts for the loops that do 
expand. The other half contracts (representing the annihilation of 
recently created pairs) and so does not contribute to the creation 
rate.  So, the one-loop factor becomes 
\begin{equation}
\frac{1}{2\pi {\cal{T}}_0} \frac{\prod \,(\lambda_n)^{-\frac{1}{2}}}
{\prod \,(\lambda'_n)^{-\frac{1}{2}}}= 
\frac{1}{2\pi {\cal{T}}_0} \frac{1}{2}\frac{i}{\sqrt{2 \pi
\epsilon}}\:\sqrt{2 \pi}R\:\frac{\prod_{\lambda> 0}
(\lambda_n)^{-\frac{1}{2}}}{\prod_{\lambda'> 0}
\,(\lambda'_n)^{-\frac{1}{2}}}=
i\frac{1}{2\pi {\cal{T}}_0}\frac{1}{2}\:\sqrt{2 \pi \epsilon}\: 
\sqrt{2 \pi}R\:.  
\label{2.23} 
\end{equation}
Written like this, the one loop factor accounts only for the 
contribution of the instanton centered in $(x_1,x_2)=(0,0)$. 
The translational invariance in the $x_1$ and $x_2$ directions
requires that one multiplies (\ref{2.23}) by the spacetime volume 
factor$\int dx_2 \int dx_1=TL $, which represents the 
spacetime region where the instanton might be localized. So, 
the correct one-loop factor is given by 
\begin{equation} 
\int [d\eta(\tau)]\exp {\biggl[}\!-\frac{1}{2}\int\!
d\tau d\tau'\,\eta_\mu(\tau)\: M_{\mu\nu}
\:\eta_\nu(\tau'){\biggr]}=i\frac{L T}{2\pi
{\cal{T}}_0}\frac{1}{2}\:\sqrt{2 \pi \epsilon}\:\sqrt{2 \pi}R\:.  
\label{2.24} 
\end{equation} 
Putting (\ref{2.24}) into (\ref{2.20}), using 
${\cal{T}}_0^2=\frac{\int d\tau
\dot{x}^2}{m^2}=\frac{(2 \pi R)^2}{m^2}\:$, 
$R=\frac{m}{\epsilon}$ and $S_0=\pi m^2/\epsilon$, 
one finally has that the soliton-antisoliton pair 
production rate per unit time and length is given 
by 
\begin{eqnarray}
 \Gamma/L = \frac{\epsilon}{2\pi}\:{\rm e}^{-\frac{\pi
m^2}{\epsilon}}\:.  
\label{2.25} 
\end{eqnarray}
We have recovered Stone's exponential factor
${\rm e}^{-\frac{\pi m^2}{\epsilon}}$ \cite{Stone}
as well as the prefactor $A=\epsilon/2\pi$
of Kiselev and Selivanov 
\cite{Kiselev1,Kiselev2} and Voloshin \cite{Voloshin1}.

Note the difference to the 4D problem of Affleck {\it et al} 
\cite{Affleck1} and Schwinger \cite{Schwinger}, who have 
found for the factor $A$ the value $(eE)^2/(2 \pi)^3$ which is 
quadratic in $eE$ and not linear, as in our case. This difference 
has to do with 
the dimensionality of the problems.

It is well known that a one-particle system in 2D can be transformed
straightforwardly to a thin line in 3D and a thin wall in 4D, where
now the mass $m$ of the soliton should be interpreted as a line
density and surface density, respectively.  In fact, a particle in
(1+1)D, as well as an infinite line in (2+1)D, can be considered as
walls as seen from within the intrinsic space dimension, justifying
the use of the name wall for any dimension.  Our calculations apply
directly to the domain wall pair creation problem in any dimension.

\vskip 2.2cm

\noindent
{\bf IV. Conclusions}
 
\vskip 3mm

The equation for the loop of radius $R$ in 2D
euclidean spacetime is given by $x^2+t_{\rm E}^2=R^2$, where we 
have put $x=x_1$ and $t_{\rm E}=x_2$. One 
can make an analytical continuation of the euclidean time 
($t_{\rm E}$) to the Minkowskian time ($t_{\rm E}=it$) and 
obtain the solution in 2D 
Minkowski spacetime 
\begin{eqnarray} 
x^2-t^2=R^2.  
\label{3.1} 
\end{eqnarray} 
At $t_{\rm E}=t=0$ the system makes a quantum 
jump and a soliton-antisoliton 
pair materializes at $x=\pm R=\pm m/\epsilon$. After 
the materialization, the 
soliton and antisoliton are accelerated, driving 
away from each other, as (\ref{3.1}) shows. To check these statements 
note first that
the energy necessary for the materialization of the pair at 
rest is $E=2m$, where $m$ is the soliton mass.
This energy comes from the conversion of false vacuum into 
true vacuum. Since $\epsilon$ is the energy difference per 
unit length between the two vacua, we conclude that 
an energy of value $E=2R \epsilon$ is released when this 
conversion occurs in the region ($2R$) within the pair.
So, the pair materialization should occur 
only when $R$ is such that the energy released is equal to 
the rest energy: $2R \epsilon=2m \Rightarrow R=m/ \epsilon$.
This value agrees with the one that has been determined 
in section III.

After the materialization the pair is accelerated so that 
its energy is now $E=2m/\sqrt{1-v^2}$. Differentiating 
(\ref{3.1}), we get the velocity $v=\sqrt{1-R^2/x^2}$. 
The energy of the pair is then given by
$E=2\frac{m}{R}|x|=\epsilon\: 2|x|$.
Notice now that $\epsilon\: 2|x|$ is the energy  
released in the conversion of false vacuum into 
true vacuum. So, after pair creation, all the energy 
released in the conversion between the two vacua is 
used to accelerate the soliton-antisoliton pair.

This discussion agrees with the interpretation of the 
process as being the false vacuum decaying to the true 
vacuum plus a creation of a soliton-antisoliton pair. 
It also justifies the presence of the interaction term 
$\epsilon\:\varepsilon_{\mu\nu}x_\nu\psi$ present in the 
covariant derivative of the proposed effective one-loop
action, (\ref{1.4}), since $\epsilon \,x$ is the energy 
released in the decay and responsible for the creation and 
acceleration of the pair.

With the proposed effective one-loop action (\ref{1.4})
we have recovered Stone's exponential factor $S_0$ 
\cite{Stone} of the pair creation rate in 
(\ref{0.1}), and the prefactor $A$ of 
Kiselev and Selivanov 
\cite{Kiselev1,Kiselev2} and Voloshin \cite{Voloshin1}.
In the proposed  effective one-loop action the soliton 
charge is treated no longer as a topological charge but as a 
Noether charge. Such an interchange between the topological 
and the Noether charges was already present in 
\cite{Coleman1,Olive}.

The problem of false vacuum decay coupled to 
gravity has been introduced in \cite{Luccia} and recently
 there has been a renewed interest in it (see, e.g., 
\cite{Garriga,Lavr}). 
With the proposed effective one-loop action 
(\ref{1.4}) we pretend to further analyse this problem.

\vskip .5cm

\section*{Acknowledgments} This work was partially funded
through project PESO/2000/PRO/40143 by the portuguese 
Funda\c c\~ao para a Ci\^encia e Tecnologia - FCT. OJCD 
also acknowledges finantial support from FCT through 
PRAXIS XXI programme. JPSL thanks Observat\'orio Nacional 
do Rio de Janeiro for hospitality.


\end{document}